\documentclass[
reprint,
 amsmath,amssymb,
 aps,
 prapplied,
]{revtex4-2}

\usepackage{graphicx}
\usepackage{dcolumn}
\usepackage{bm}
\usepackage{subcaption}
\usepackage{float}
\usepackage{hyperref}
\usepackage{etoolbox}
\usepackage{orcidlink}

\newcommand{\swt}{\mathrm{\zeta_{t}}}

\begin{document}


\title{Field-free and Sub-ns Magnetization Reversal in Synthetic Anti-ferromagnets using Sequential Spin-Orbit Torque and Spin-Transfer Torque Pulses}

\author{Subham Das\,\orcidlink{0009-0002-7555-6903}}
\email{dassubham@iitgn.ac.in}

\author{Naveen Sisodia\,\orcidlink{0000-0002-8636-5400}}
\email{naveen.sisodia@iitgn.ac.in}
\affiliation{
 Department of Physics, Indian Institute of Technology Gandhinagar,\\
 Gandhinagar, Gujarat, India - 382355
}

\date{\today}

\begin{abstract}
Single free-layer Magnetic Tunnel Junctions (MTJs) with a perpendicular easy axis driven by spin currents have demonstrated fast, deterministic switching, non-volatility, and low-power memory capabilities when integrated with modern nanoelectronic devices. In this article, we numerically investigate sub-ns and field-free switching characteristics of a p-MTJ with a Synthetic Antiferromagnet (SAF) based free layer structure using sequential sub-ns current pulses of Spin Orbit Torque (SOT) and Spin Transfer Torque (STT). While the SOT pulse applied to one layer decreases the transition time by bringing the coupled layers into an in-plane configuration, the subsequent STT pulse on the system determines the switching direction. Various pulse protocols have been investigated for both SOT and STT current amplitude and pulse widths, while also considering the effect of the Reference Layer (RL). While significant regions of deterministic switching events are observed, limitations in SOT, STT parameters are further investigated in presence of thermal fluctuations.

\end{abstract}

\maketitle

\section*{\label{sec:intro}Introduction}
Magnetic Random Access Memories (MRAMs) are increasingly considered a promising alternative to conventional CMOS-based memories due to the growing demand for energy-efficient computing across applications ranging from large-scale sensor networks to neuromorphic computing\cite{mramsensor,mramneuro}. Magnetic Tunnel Junctions (MTJs), which are the fundamental building blocks of MRAMs, offer non-volatile storage with bit states that can be manipulated by electric currents/voltages, and are directly compatible with CMOS processes\cite{mramcmos}. These MTJs are primarily composed of a trilayer structure with two magnetic layers separated by an oxide barrier. The magnetic states with parallel (P) and antiparallel (AP) configurations of the two magnetic layers give rise to different resistance states via Tunnel Magnetoresistance (TMR). One of the magnetic layers in this structure, called the free layer (FL), is actively manipulated using spin currents to switch between the P and AP states. In particular, MTJs with FL having a perpendicular anisotropy (p-MTJs) provide higher storage density, high thermal stability and lower switching currents compared to MTJs with in-plane FL, enabling ultrafast and energy-efficient read and write operations\cite{ip-oop-mtj,ip-oop-mtj2}. The spin currents required for switching are typically generated in MTJs using Spin Transfer Torque (STT) or Spin Orbit Torque (SOT) mechanisms\cite{SLONCZEWSKI1996L1,Berger1996, Miron2011}. 

STT-MRAMs consist of two-terminal MTJs with identical read and write paths, which imposes limitations in high-speed applications as high write currents can potentially damage the oxide barrier. In addition, STT-MTJs also suffer from high incubation times for switching, as the initial torque on the magnetization is vanishingly small, leading to slower switching speeds. Three-terminal SOT-driven p-MTJs provide lower incubation times with separate read and write paths due to the utilization of spin currents generated by Spin Hall Effect (SHE)\cite{spinhall}. 
However, achieving deterministic switching in p-MTJs using SOT is challenging and requires external fields/currents to break the symmetry of SOT torque. This asymmetry can be introduced using external fields\cite{singleSOT+field}, stray fields\cite{SOTstray}, exchange coupling to an antiferromagnet\cite{Fukami2016,Oh2016}, geometrical asymmetry\cite{Safeer2016,Yu2014} or Voltage Controlled Magnetic Anisotropy (VCMA)\cite{Zhang2016,ZhangXian2015} in the system. A combination of SOT and STT has also been implemented, where SOT reduces the incubation and transition times, thereby reducing the STT write currents, while the deterministic switching is carried out using STT.\cite{SOTSTTsingle1,Pathak2020,SOTSTTsingle2}. A careful tuning of current polarity for STT and SOT is required in such cases, as SOT can assist or counteract STT, depending on the relative polarities of the two, when they are applied simultaneously\cite{IP_SOT+-STT}. In addition, simultaneous injection of both SOT and STT pulses also requires careful engineering design due to complex current paths arising from the application of multiple voltages at different terminals.

Recently, Synthetic Antiferromagnetic (SAF) structures have gained considerable attention due to their faster dynamics, higher thermal stability, decreased stray fields, and increased robustness against external fluctuations\cite{SAFverify1,SAFverify2}. SAF are trilayer structures with two magnetic layers coupled antiferromagnetically to each other via a non-magnetic (NM) spacer layer. The composition of the reference layer in the MTJ structure already utilizes such SAF structures in mitigating the dipolar field arising from RL\cite{singleSOT+SAF,Grimaldi2020}. SAF structures have also been introduced in the FL layer of the MTJ, replacing the single ferromagnetic free layer in the conventional MTJ (RL/Oxide/\textbf{FM}) with a trilayer (RL/oxide/\textbf{FM/NM/FM}) Synthetic anti-ferromagnet (SAF) structures.\cite{SAFSOTgen,SAF+single,SAF-VCMA-VCEC-SOT,SAFstrain,pan2025field,SAFstochastic,SAFSOTuncomp} 

In this work, we numerically study the magnetization reversal dynamics of a p-MTJ with a compensated SAF as the free layer, by using sequential SOT and STT sub-ns pulses. We study the effects of pulse amplitude and width for both SOT and STT pulses on the dynamics of switching in the SAF free layer and present comprehensive phase maps showing the regions of deterministic switching. We characterize the switching processes by calculating the switching times and threshold current densities required to perform complete switching between different magnetization states and present an optimization strategy to obtain energy-efficient sub-ns switching. The article is structured as follows: Sec.~\ref{sec:system} describes the device structure and outlines the simulation methodology. Limitations in switching with Pure STT and Pure SOT are discussed in Sec.~\ref{sec:pureSTTall}\&~\ref{sec:pureSOT} respectively. Magnetization reversal dynamics using the proposed protocol is shown in Sec.~\ref{sec:SOT+STT}. Sec.~\ref{sec:SOTmaps}\&~\ref{sec:STTmaps} show the phasemaps of switching time for both SOT and STT current densities and pulse-widths, including the effect of the Reference Layer. The results are summarised in the Conclusion section along with a discussion on the limitations and future scope of the study.

\section{\label{sec:system} Hybrid SAF Device Model}
\begin{figure}[H]
    \begin{minipage}{1.0\columnwidth}
        \centering
       \includegraphics[width=\columnwidth]{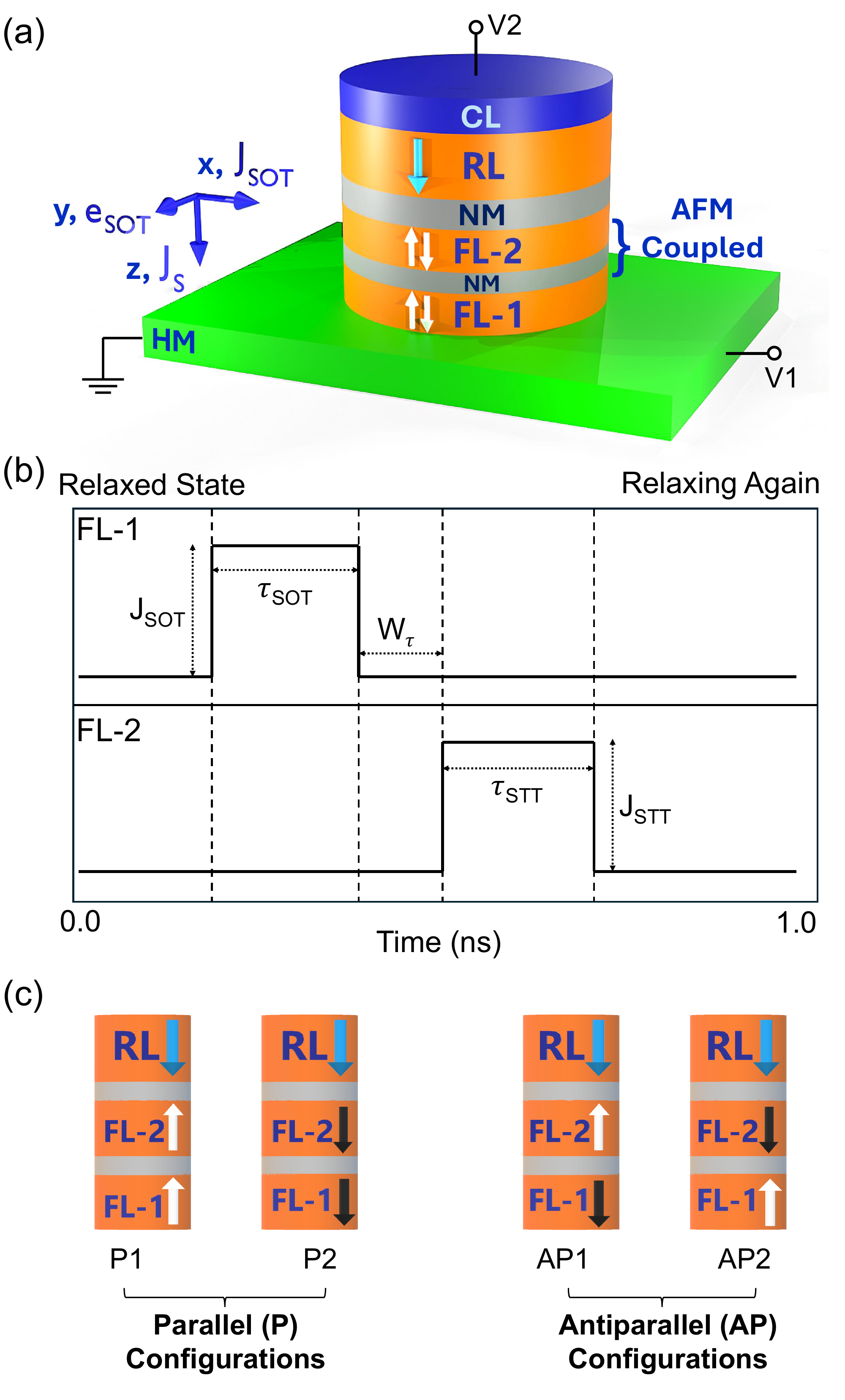}
        \caption{\label{fig:1diagram} (a) A schematic of a three-terminal SAF free-layer based p-MTJ device. (b) SOT and STT pulse profile with all the tunable parameters. $J_{\rm STT}$ and $J_{\rm SOT}$ represent the amplitude of STT and SOT currents, respectively, while $\tau_{\rm STT}$ and $\tau_{\rm SOT}$ represent the corresponding current pulse widths. The parameter $W_{\tau}$ represents the delay between the SOT and STT pulses. $W_{\tau}$ is taken to be zero for this work.
        (c) Four possible magnetization states of the system are shown, with parallel configurations (P1 and P2) on the left and anti-parallel configurations (AP1 and AP2) on the right.
        }
    \end{minipage}
\end{figure}

Fig.~\ref{fig:1diagram}(a) shows the simulated structure of the compensated SAF-based p-MTJ device with a diameter of $\rm 20~nm$. The top layer (blue) is the capping layer (CL), followed by a ferromagnetic reference layer (RL). RL is followed by an oxide barrier separating it from the SAF-free layer structure. The SAF structure, comprising FL-1 and FL-2 layers coupled antiferromagnetically via the non-magnetic (NM) layer, constitutes the free layer of the MTJ. Both FL-1 and FL-2 have PMA with an out-of-plane easy axis. Finally, FL-1 is in contact with a heavy-metal layer (HM), which applies Spin-Orbit Torque (SOT) to FL-1 due to the injection of spin-polarized current generated via Spin Hall Effect (SHE) at FL-1/HM interface. Similarly, charge current injected via contacts V1 and V2 is converted to a spin-polarized current due to RL and applies Spin-Transfer Torque (STT) on FL-2. The effect of STT torque on FL-1 is assumed to be small and hence has not been included in the simulations. 

The magnetization dynamics of the ferromagnetic free layers are investigated in a micromagnetic framework implemented using Mumax3\cite{Vansteenkiste2014} package. The Landau-Lifshitz-Gilbert (LLG) equation\cite{landau_lifshitz,gilbertDamp} governing the classical magnetization dynamics is solved using the Dormand-Prince solver\cite{DORMAND_prince}, along with the additional SOT and STT terms (Eq.~(\ref{eq:LLG1}),~(\ref{eq:LLG2})).

In our simulations, the effective field term includes both inter-layer and intra-layer exchange field terms as well as magneto-static and anisotropic field terms. Only the damping-like torque terms for SOT and STT are incorporated in the simulations(Eq.~(\ref{eq:LLGSOT}), ~(\ref{eq:LLGSTT})).
\begin{eqnarray}
    \frac{d\hat{m}_1}{dt}&=&-\frac{\gamma}{1+\alpha^2}(\hat{m}_1 \times \overrightarrow{B}_{eff,1})\nonumber\\
    & &-\frac{\gamma\alpha}{1+\alpha^2}[\hat{m}_1 \times(\hat{m}_1 \times \overrightarrow{B}_{eff,1})]+\tau_{SOT}
    \label{eq:LLG1}
\end{eqnarray}
\begin{eqnarray}
    \frac{d\hat{m}_2}{dt}&=&-\frac{\gamma}{1+\alpha^2}(\hat{m}_2 \times \overrightarrow{B}_{eff,2})\nonumber\\
    & &-\frac{\gamma\alpha}{1+\alpha^2}[\hat{m}_2 \times(\hat{m}_2 \times \overrightarrow{B}_{eff,2})]+\tau_{STT}
    \label{eq:LLG2}
\end{eqnarray}
\begin{eqnarray}
    \overrightarrow{B}_{IEC,1}=2J_{ex}(\hat{m}_2-\hat{m}_1)
    \label{eq:IEC1}
\end{eqnarray}
\begin{eqnarray}
    \overrightarrow{B}_{IEC,2}=2J_{ex}(\hat{m}_1-\hat{m}_2)
    \label{eq:IEC2}
\end{eqnarray}
\begin{eqnarray}
    \overrightarrow{\tau}_{SOT}&=&-\frac{\gamma}{1+\alpha^2}\beta_{SOT}[\hat{m}_1 \times(\hat{m}_1 \times \hat{\sigma})]\nonumber\\
    & &+\frac{\gamma \alpha}{1+\alpha^2}\beta_{SOT}(\hat{m}_1 \times \hat{\sigma})
    \label{eq:LLGSOT}
\end{eqnarray}
\begin{eqnarray}
    \overrightarrow{\tau}_{STT}&=&-\frac{\gamma}{1+\alpha^2}\beta_{STT}[\hat{m}_2 \times(\hat{m}_2 \times \hat{m}_P)]\nonumber\\
    & &+\frac{\gamma \alpha}{1+\alpha^2}\beta_{STT}(\hat{m}_2 \times \hat{m}_P)
    \label{eq:LLGSTT}
\end{eqnarray}
\begin{equation}
    \beta_{SOT}=\frac{\hbar J_{z,SOT}P_{SOT}}{2M_Sed};\ \ \beta_{STT}=\frac{\hbar J_{z,STT}P_{STT}}{2M_Sed}
    \label{eq:betaST}
\end{equation}
Here, ${\overrightarrow{m}_1}$, ${\overrightarrow{m}_2}$ represent the normalized magnetization of FL-1 and FL-2, respectively. ${\overrightarrow{\sigma}}$, and ${\overrightarrow{m}_P}$ are the spin polarization directions for SOT and STT, respectively. $\overrightarrow{B}_{eff,1}$ and $\overrightarrow{B}_{eff,2}$ are the effective field terms for FL-1 and FL-2, respectively, that contain magnetostatic, anisotropic, intra-layer, and inter-layer exchange field terms. The inter-layer exchange terms added in the effective field terms, are shown in equations (\ref{eq:IEC1}) and (\ref{eq:IEC2})\cite{IEC1,IEC2}. $J_{\rm ex}$ is the inter-layer exchange constant, where its negative value determines the antiferromagnetic exchange coupling between the two free layers. The z-components of the current densities associated with SOT and STT are denoted by \( J_{\rm z,SOT} \) and \( J_{\rm z,STT} \), respectively; throughout this article, they are referred to as \( J_{\rm SOT} \) and \( J_{\rm STT} \). The direction of spin polarization for SOT is along $\hat{y}$, as determined by the polarity of the flow of charge current in the HM layer (${|\overrightarrow{\sigma}|.\hat{\sigma}=\overrightarrow{J}_{\rm C,SOT}\times \overrightarrow{J}_{S,SOT}}$). The polarization of STT current is along $\hat{z}$ direction as indicated in Fig.~\ref{fig:1diagram}(a). The cell size for both free layers is fixed at ($\mathrm{0.3125\times0.3125\times1}$) ${\rm nm^{3}}$.
\begin{table}[H]
\caption{\label{tab:table1param}%
Material parameters for the simulated device geometry}
\begin{ruledtabular}
\begin{tabular}{ccc}
\textrm{Parameters} &
\textrm{Description} &
\textrm{Value} \\
\colrule
$M_S$ & Saturation magnetization & $1\,\mathrm{MA/m}$ \cite{Carpentieri2018} \\
$K_{u_{1}}$ & 1st order anisotropy coefficient & $0.8\,\mathrm{MJ/m^3}$ \cite{Carpentieri2018} \\
$d$ & Thickness of Free Layers & $1\,\mathrm{nm}$ \cite{Carpentieri2018} \\
$D$ & Diameter of Free Layers & $20\,\mathrm{nm}$ \\
$\alpha$ & Damping constant & $0.1$ \\
$A_{ex}$ & Intra-layer exchange constant & $16\,\mathrm{pJ/m}$ \\
$J_{ex}$ & Inter-layer exchange constant & $-0.2\,\mathrm{T}$ \\
$P_{STT}$ & Spin Transfer Torque Efficiency & $0.66$ \cite{Carpentieri2018} \\
$P_{SOT}$ & Spin-Orbit Torque Efficiency & $0.12$ \cite{Pathak2020} \\
\end{tabular}
\end{ruledtabular}
\end{table}

The successive SOT-STT pulse protocol used for obtaining the primary results of this paper is shown in Fig.~\ref{fig:1diagram}(b). The top inset shows the pulse profile of the SOT pulse, applied to the FL-1 layer, and the bottom inset shows the pulse profile of the STT pulse acting on the FL-2 layer. The tunable parameters are the current densities, $(\mathrm{J_{SOT}, J_{STT}})$ and the pulse-widths, $(\mathrm{\tau_{SOT}, \tau_{STT}})$ of SOT and STT currents. The pulse-gap, $\mathrm{W_{\tau}}$ between SOT and STT pulses is taken to be zero for all simulations. The material parameters used in the simulations are similar to the CoFeB Free layer based MTJ system used by Carpentieri \textit{et al.}\cite{Carpentieri2018} and are given in Table~\ref{tab:table1param}.

\section{\label{sec:disc}Results and discussions}

The SAF-pMTJ device offers four possible magnetization states as shown in Fig.\ref{fig:1diagram}(c), with two parallel (P) and two antiparallel (AP) states. For our work, we focus mainly on the switching between the two low-energy AP-states (AP1 and AP2), so that the net magnetization remains zero for both states, creating minimal stray field while maintaining robustness against external fields (compensated SAF). The resistance change between the two states is mainly due to the parallel or antiparallel alignment of FL-2 relative to the RL. It may be noted that while the energy of both AP-states is the same for an isolated trilayer-SAF system, it does not remain equal in the presence of the RL stray field. The effect of the RL stray field on switching characteristics is discussed separately in Appendix ~\ref{sec:RLeff}. 

To understand the basic switching operation in the SAF-pMTJ device using spin currents, we first examine the effects of pure STT and pure SOT pulses separately in Sec.~\ref{sec:pureSTTall} and Sec.~\ref{sec:pureSOT}. In Sec .~\ref {sec:SOT+STT}, we show the magnetization switching using our proposed successive SOT-STT protocol and compare it with the results obtained by applying pure STT and SOT pulses.

\subsection{\label{sec:pureSTTall}Switching with Pure STT}

\begin{figure}[H]
    \begin{minipage}{1.0\columnwidth}
        \centering
       \includegraphics[width=\columnwidth]{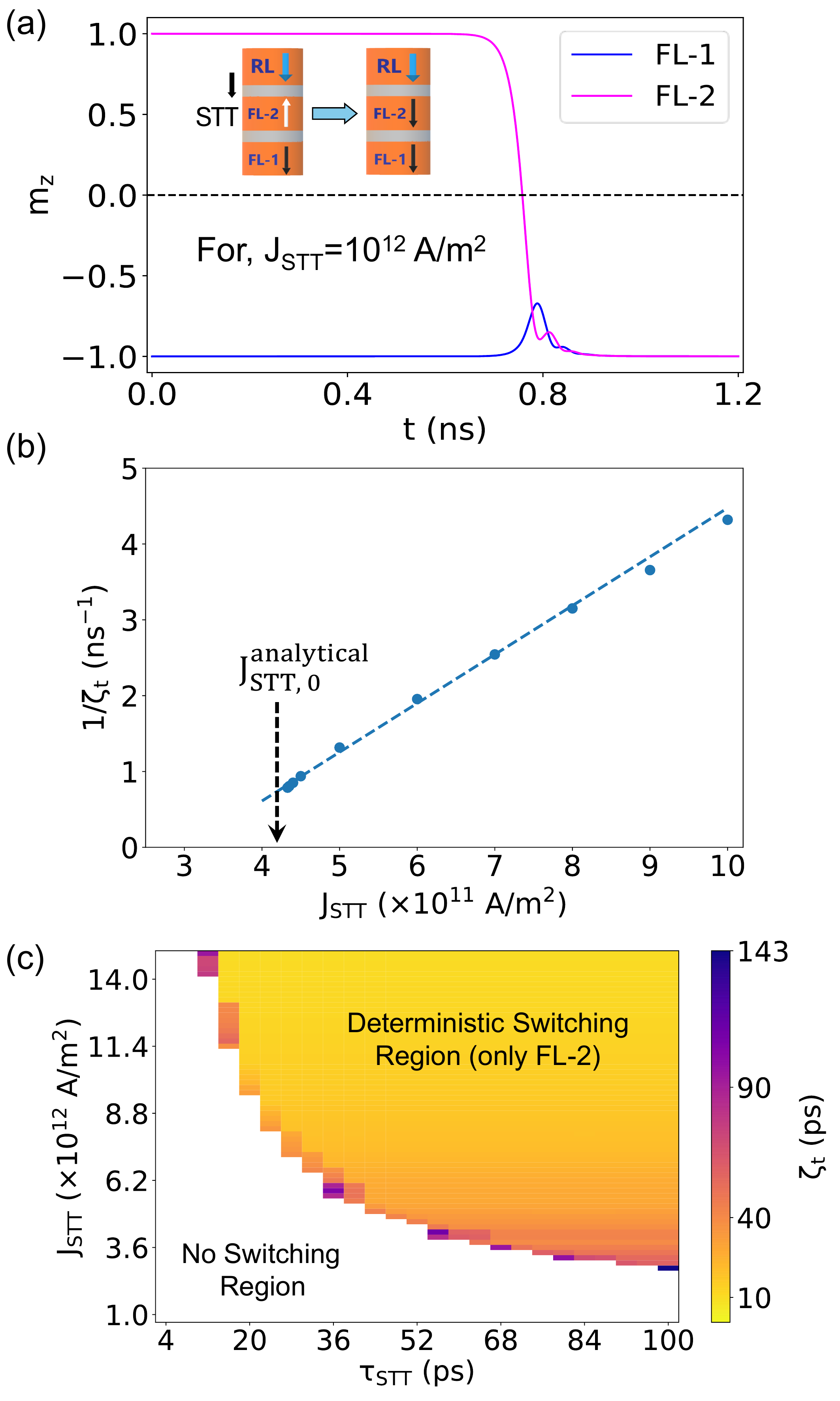}
        \caption{\label{fig:2abc} (a) Change in the z-component of magnetization, $m_z$, as a function of time with the application of continuous STT current of amplitude, $\mathrm{J_{STT} = 10^{12}\,A/m^2}$. (b) Inverse of Switching time for FL-2 ($1/\swt$) vs. $J_{STT}$. The extrapolated threshold current density is approximately $3\times 10^{11}\,A/m^2$. (c) Switching Time ($\swt$) Phasemap for various values of STT current amplitudes ($\mathrm{J_{STT}}$) and pulsewidths ($\mathrm{\tau_{STT}}$). Color represents the switching time, $\swt$. }
    \end{minipage}
\end{figure}

For switching using pure STT, the current pulse is injected through contacts V1 and V2, and applies an STT torque on FL-2. We assume that the FL-1 experiences no STT torque and that its dynamics are mainly due to antiferromagnetic coupling with FL-2. We first apply a continuous ($\mathrm{\tau_{STT}=\infty}$) current pulse with $\mathrm{J_{STT}=10^{12}\,A/m^2}$ and show the time-varying z-component of the magnetization ($\mathrm{m_z}$) for both FL-1 and FL-2 in Fig.~\ref{fig:2abc}(a). 
Here, the switching is only observed in FL-2 at $\mathrm{t\approx0.8~ns}$ while the FL-1 remains in the same state, leading to an $\mathrm{AP1\rightarrow P2}$ transition, as shown in the inset of Fig.~\ref{fig:2abc}(a). The magnetization state inside both FM layers remains nearly uniform throughout the switching. The transition involves a long incubation time of $\mathrm{\sim 0.7~ns}$ for the FL-2 layer, followed by a rapid change of FL-2 magnetization from $+z$ to $-z$-direction. The Switching Time ($\swt$) is defined as the total time taken for the transition from $\mathrm{m_z=+0.99}$ to $\mathrm{m_z=-0.99}$. During the rotation of FL-2 magnetization, a change in FL-1 magnetization is also observed at $\rm 0.8~ns$ due to the torque from the inter-layer exchange coupling (IEC). However, for the chosen parameters, this torque is insufficient to overcome the anisotropy energy barrier in FL-1. While the strength of this torque can be increased by introducing stronger exchange coupling, such an increase will also lead to a significant increase in the switching current and is thus not considered to be an optimal pathway for $\mathrm{AP1\rightarrow AP2}$ switching. An increase in the STT current density, while keeping the inter-layer exchange strength fixed, shows qualitatively the same behaviour as seen in Fig.~\ref{fig:2abc}(a) with a reduced switching time. Figure~\ref{fig:2abc}(b) shows the variation of the inverse of the switching time with applied STT current density. The data points are linearly fitted, and the intercept of the fit on the $\mathrm{J_{STT}}$-axis represents the threshold STT current density ($\mathrm{J_{STT,0}}$), at which the switching time becomes infinite. From the fit, $\mathrm{J_{STT,0}}$ was calculated to be $\sim\mathrm{3\times10^{11}\,A/m^2}$. We also calculate this threshold current density analytically using a coupled-macrospin model. We assume a nearly uniform magnetization profile for both FM layers in Eqs.~(\ref{eq:LLG1}) and (\ref{eq:LLG2}). The $\tau_{\rm SOT}$ term is set to zero. Stereographic projection is used to convert the magnetization vector to a complex variable, one for each FM layer. The coupled equations are then linearized and solved for the switching event. Details of the calculations are given in Appendix~\ref{sec:analytSTTcont}. For our parameters, $\mathrm{J^{analytical}_{STT,0}}$ is calculated to be $\sim\mathrm{4.18\times10^{11}\,A/m^2}$.

To understand the switching dynamics under pulsed current, we now apply STT current pulses with varying amplitudes ($\mathrm{J_{STT}}$) and finite pulse width ($\mathrm{\tau_{STT}}$). Fig.~\ref{fig:2abc}(c) shows region of switching as a function of varying $\mathrm{J_{STT}}$ and $\mathrm{\tau_{STT}}$ with the color representing the switching time, $\mathrm{\swt}$, for FL-2. Similar to Fig.~\ref{fig:2abc}(a), the switching is observed only in FL-2; however, we observe that the switching in this case occurs via a non-uniform vortex-like state. For a fixed $\tau_{STT}$, $\mathrm{\swt}$ decreases with increasing current density and pulse width, consistent with the expected behaviour. Additionally, for any fixed $\mathrm{\tau_{STT}}$, a threshold $\mathrm{J_{STT}}$ exists below which no switching is observed. Similarly, for a fixed $\mathrm{J_{STT}}$, a minimum $\mathrm{\tau_{STT}}$ is required for FL-2 magnetization to respond and eventually switch. It is important to note that the $\mathrm{J_{STT}}$ values in Fig.~\ref{fig:2abc}(c) are roughly one order of magnitude higher than the $\mathrm{J_{STT}}$ values in Fig.~\ref{fig:2abc}(a), where the current was applied continuously. 
Therefore, the faster sub-ns switching comes at the cost of using higher $\mathrm{J_{STT}}$. This requirement for high $\mathrm{J_{STT}}$ for shorter current pulses can be attributed directly to the large incubation time for pure STT-driven switching, since the torque acting on the magnetization aligned along $z$-axis is vanishingly small for the STT pulse with $+z$ polarization. 

\subsection{\label{sec:pureSOT}Switching with Pure SOT Pulse}
The pure SOT-driven switching is studied by injecting current pulses through contacts V1 and ground (Fig.~\ref{fig:1diagram}(a)). The spin current with polarization along $\pm y$, generated via the Spin Hall Effect (SHE) on FM-1/HM interface, applies a torque directly on FL-1, while FL-2 only experiences a torque due to interlayer exchange coupling.
\begin{figure}[H]
    \begin{minipage}{1.0\columnwidth}
        \centering
       \includegraphics[width=\columnwidth]{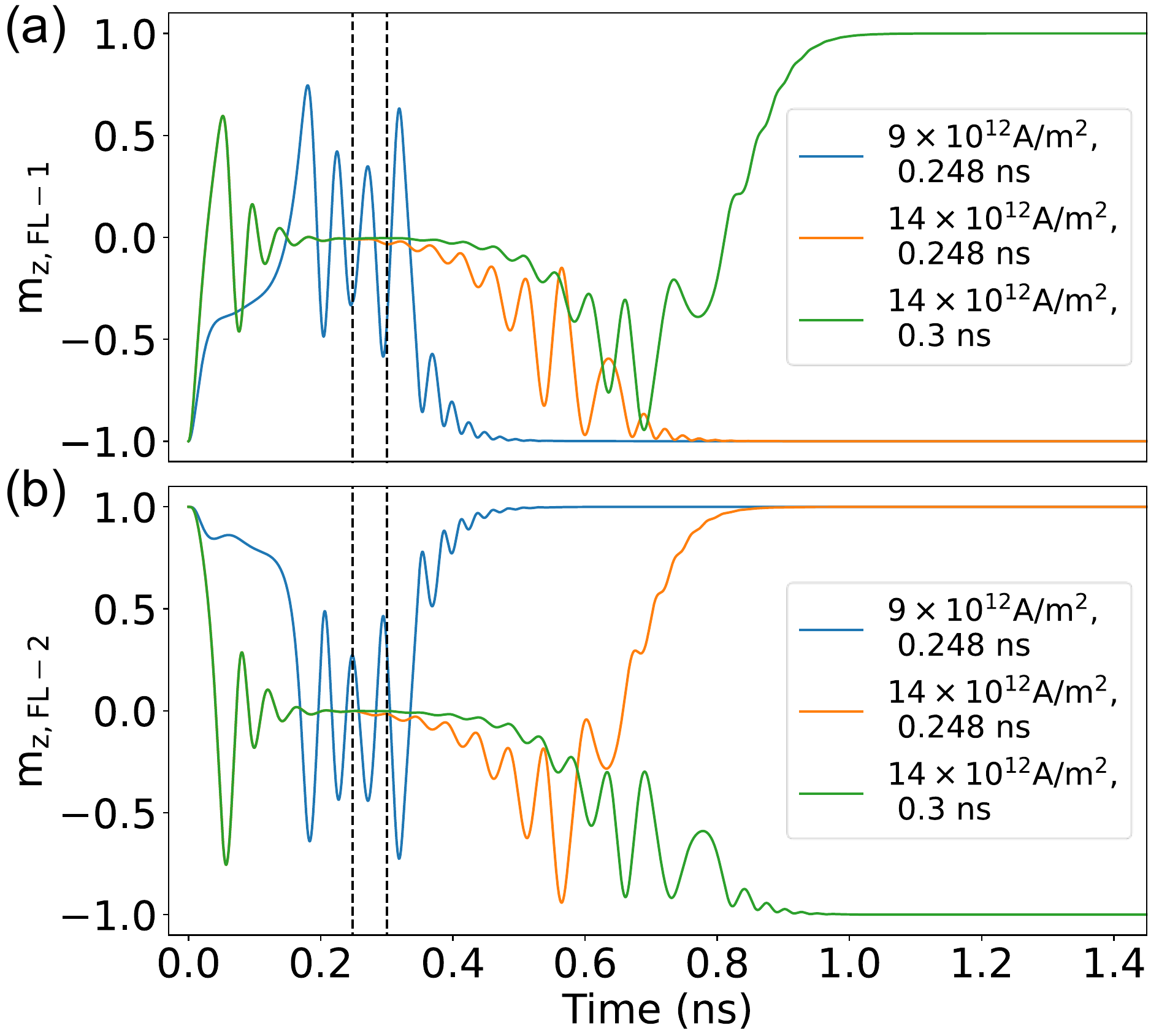}
        \caption{$\mathrm{m_z}$ components of (a) FL-1, (b) FL-2, vs. time for various values of $\mathrm{J_{SOT}}$ and $\mathrm{\tau_{SOT}}$ when only SOT is applied on FL-1.}
        \label{fig:SOTstoc}
    \end{minipage}
\end{figure}
Figure~\ref{fig:SOTstoc}(a) shows the $\rm m_z$ components of FL-1 and FL-2 for three different combinations of SOT pulse amplitude $\mathrm{J_{SOT}}$ and width $\mathrm{\tau_{SOT}}$. For all three cases, in contrast to the pure-STT case, where a large incubation time was seen before the beginning of the $\rm m_z$ transition, for pure-SOT cases, the transition begins immediately with magnetization rotating from out-of-plane (OOP) to in-plane (IP) state with a precessional motion. For a smaller $\mathrm{J_{SOT}}$ value of $\rm 9\times 10^{12}~A/m^2$, the magnetization oscillations have a very large amplitude. For larger $\mathrm{J_{SOT}}$ value, this precession quickly decays and pushes the magnetization into a pure IP state. This is seen for the two cases with $\mathrm{J_{SOT}}=14\times 10^{12}~A/m^2$ where $\rm m_{z, FL1}=m_{z, FL2}\approx 0$ at $\rm \sim 0.2~ns$. In all cases, once the SOT pulse ends, magnetization starts to relax back to either AP1 ($\mathrm{J_{SOT}}=14\times 10^{12}~A/m^2$, $\mathrm{\tau_{SOT}}=0.248~\rm ns$) or AP2 ($\mathrm{J_{SOT}}=14\times 10^{12}~A/m^2$, $\mathrm{\tau_{SOT}}=0.3~\rm ns$) state, leading to a precessional ``toggle-like'' switching behaviour similar to what has been observed for single-layer MTJs\cite{ZhangXian2015}.

\begin{figure}[H]
    \begin{minipage}{1.0\columnwidth}
        \centering
       \includegraphics[width=\columnwidth]{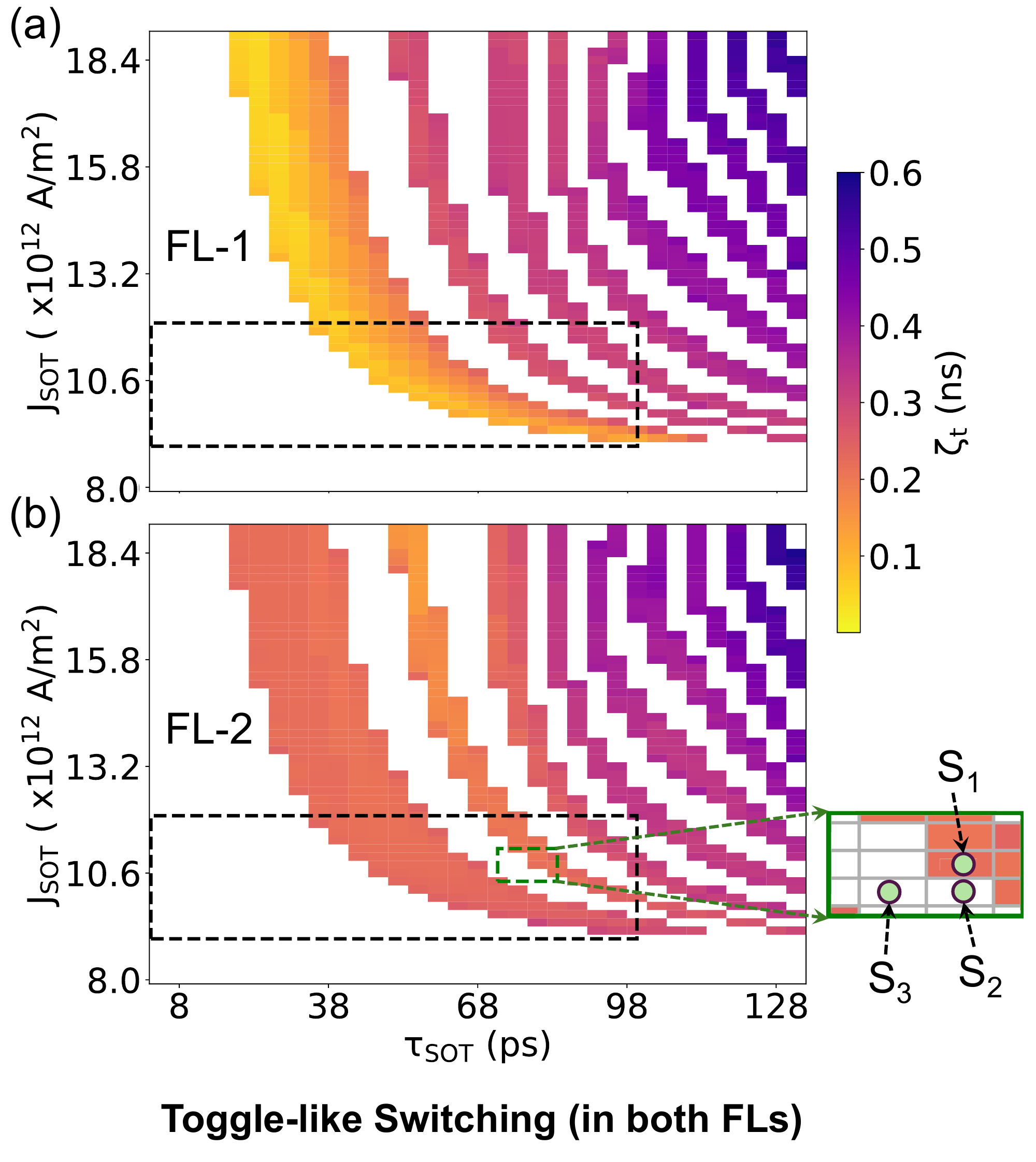}
        \caption{Switching Time ($\mathrm{\zeta_{t}}$) phasemap for various values of SOT current amplitudes ($\mathrm{J_{SOT}}$) and pulsewidths ($\mathrm{\tau_{SOT}}$); (a) for FL-1, (b) for FL-2. The black dashed rectangles show the region of interest for SOT+STT driven switching discussed in Sec.~\ref{sec:SOT+STT}. The outset of (b) shows a magnified view of the phasemap within the green dashed rectangle, highlighting three neighboring points ($\mathrm{S_1,~S_2,~and~S_3}$) selected to obtain the phasemaps for STT parameters (Fig.~\ref{fig:sotstt_map_vary_STT}).}
    \label{fig:puresotpulse}
    \end{minipage}
\end{figure}

To further explore SOT-driven switching, we systematically vary the SOT current pulse amplitude ($\mathrm{J_{SOT}}$) and pulse width ($\mathrm{\tau_{SOT}}$) and calculate the switching phase-maps for FL-1 and FL-2 as shown in Fig.~\ref{fig:puresotpulse}(a) and (b), respectively. The color represents the switching time, $\swt$, for individual FL-1 and FL-2 layers, with the white region representing unsuccessful switching events. For all combinations of $\mathrm{J_{SOT}}$ and $\mathrm{\tau_{SOT}}$ where switching took place, $\rm AP1\rightarrow AP2$ transition was obtained via uniform states, with no intermediate vortex state. The switching with pure SOT mainly requires the magnetization of both layers, which are initially aligned along $\pm z$-direction ($\rm m_z=\pm1$) to cross the equator ($\rm m_z=0$). The minimum energy required to reach this condition gives the lower threshold $\mathrm{J_{SOT}}-\mathrm{\tau_{SOT}}$ curve. High $\mathrm{J_{SOT}}$ with low $\mathrm{\tau_{SOT}}$ leads to faster switching as the strong SOT torque quickly brings the magnetization to the IP state as seen in Fig.~\ref{fig:puresotpulse}(a) and (b).

Importantly, for fixed $\mathrm{J_{SOT}}$, an increase in $\mathrm{\tau_{SOT}}$ above the minimum required threshold delays the switching process, leading to higher switching time, as the magnetization has already reached the IP state and will start to switch only after SOT is switched off. It may also be noted from Figs.~\ref{fig:puresotpulse}(a) and (b) that the switching time for both FL layers is not equal. This is due to the unequal competing torques acting on each layer, which leads to an additional phase difference between FL1 and FL2 as they move away from the pure antiparallel alignment. The overall switching time for the device can thus be defined as the maximum of the switching time for FL-1 and FL-2, i.e., for any set of current pulse parameters $\zeta_{\rm t, max}=max(\zeta_{t, FL-1},\zeta_{t, FL-2})$. 

\subsection{\label{sec:SOT+STT}Switching with Successive SOT and STT Pulses}
Although the simultaneous switching of both FL-1 and FL-2 was achieved in the previous section using pure SOT-pulse, the switching remained toggle-like with only narrow bands of optimal $\mathrm{J_{SOT}}$, $\mathrm{\tau_{SOT}}$ values giving $\rm AP1\rightarrow AP2$ transition. To obtain a larger range of switching in $\mathrm{J_{SOT}}-\mathrm{\tau_{SOT}}$  colormap, we further optimize the current protocol and add a STT pulse to FL-2 layer immediately after the end of the SOT pulse which was acting on FL-1 layer. The delay time defined in Fig.~\ref{fig:1diagram}(b) is $\rm W_{\tau}=0~ns$, which is kept the same throughout this article. 

\begin{figure}[H]
    \begin{minipage}{1.0\columnwidth}
        \centering
       \includegraphics[width=\columnwidth]{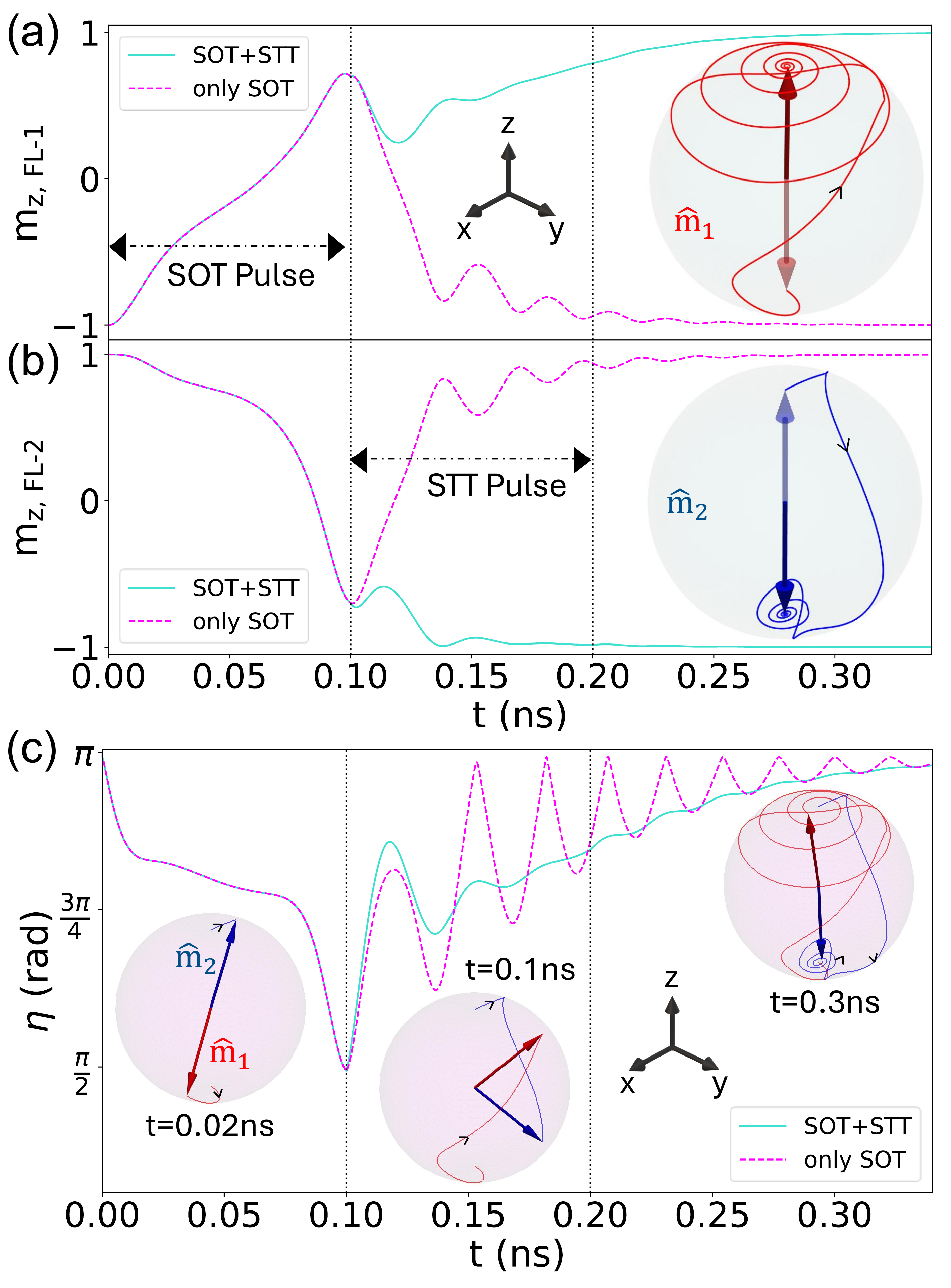}
        \caption{$\mathrm{m_z}$ components of (a) FL-1, (b) FL-2, vs. time for pure SOT in FL-1 (dashed line) and sequential SOT+STT case (solid line). The parameters for SOT, $\mathrm{J_{SOT}=9.6\times 10^{12}\,A/m^2}$ and$\mathrm{\tau_{SOT}=0.1\,ns}$, are same in both cases. The parameters for STT pulse in SOT+STT case are $\mathrm{J_{STT}=1\times 10^{12}\,A/m^2}$, and $\mathrm{\tau_{STT}=0.1\,ns}$. The insets in (a) and (b) show the trajectory of average magnetization mapped onto a unit sphere for FL1 and FL2, respectively. The red and blue arrows represent the initial and final magnetization state of each FM layer. (c) shows the angle between FL-1 and FL-2 as a function of time. The insets show the magnetization state of both FL-1 and FL-2 mapped simultaneously on a unit sphere at three different instants of time (0.02, 0.1, and 0.3ns).
        }
        \label{fig:sotstt_mz_t}
    \end{minipage}
\end{figure}

Figure~\ref{fig:sotstt_mz_t}(a) and (b) show the magnetization switching for FL-1 and FL-2, respectively, using a SOT pulse with $\mathrm{J_{SOT}=9.6\times 10^{12}\,A/m^2}$ and $\mathrm{\tau_{SOT}=0.1\,ns}$ immediately followed by an STT pulse with $\mathrm{J_{STT}=1\times 10^{12}\,A/m^2}$, and $\mathrm{\tau_{STT}=0.1\,ns}$. For comparison, the $\rm m_z$ variation with time for the pure SOT case with the same SOT pulse parameters (and STT=0) is also shown alongside. It can be seen that until the end of the SOT pulse, the magnetizations in both SOT and SOT+STT cases follow the same trajectory, as expected. However, once the STT pulse is applied, the magnetization is tilted away from the projected trajectory of the pure SOT case and is forced to transition to the AP2 state. This behaviour arises because once the SOT is turned off, the STT drives the FL-2 magnetization toward the desired direction, while the FL-1 magnetization relaxes under the influence of the IEC, leading to a $\rm AP1\rightarrow AP2$ transition. It is noted that no intermediate non-uniform or vortex-like states are observed even after the injection of additional STT pulse, in contrast to the behaviour for pure STT-driven switching. The insets of Figs.~\ref{fig:sotstt_mz_t}(a) and (b) show the FL-1 and FL-2 trajectory of average magnetisation, respectively, for the SOT+STT case mapped to a unit sphere. The arrows represent the initial and final average magnetizations.

Similar to the pure SOT case, the FL-1 and FL-2 magnetizations for SOT+STT case do not remain exactly anti-parallel during the switching process due to multiple competing interactions between SOT, STT, IEC and magnetic anisotropy. Figure ~\ref{fig:sotstt_mz_t}(c) shows the variation of angle, $\eta$, between FL-1 and FL-2, showing intermediate canted magnetization states with $\eta$ reaching even close to $\pi/2$. However, the canted precession eventually decays to the desired AP2 state. The insets of Fig.~\ref{fig:sotstt_mz_t}(c) show the magnetization states of both FL-1 and FL-2 mapped to a unit sphere at different instants of time.

\subsubsection{\label{sec:SOTmaps}Switching Phasemaps for SOT parameters}
\begin{figure}[H]
    \begin{minipage}{1.0\columnwidth}
        \centering
    \includegraphics[width=\columnwidth]{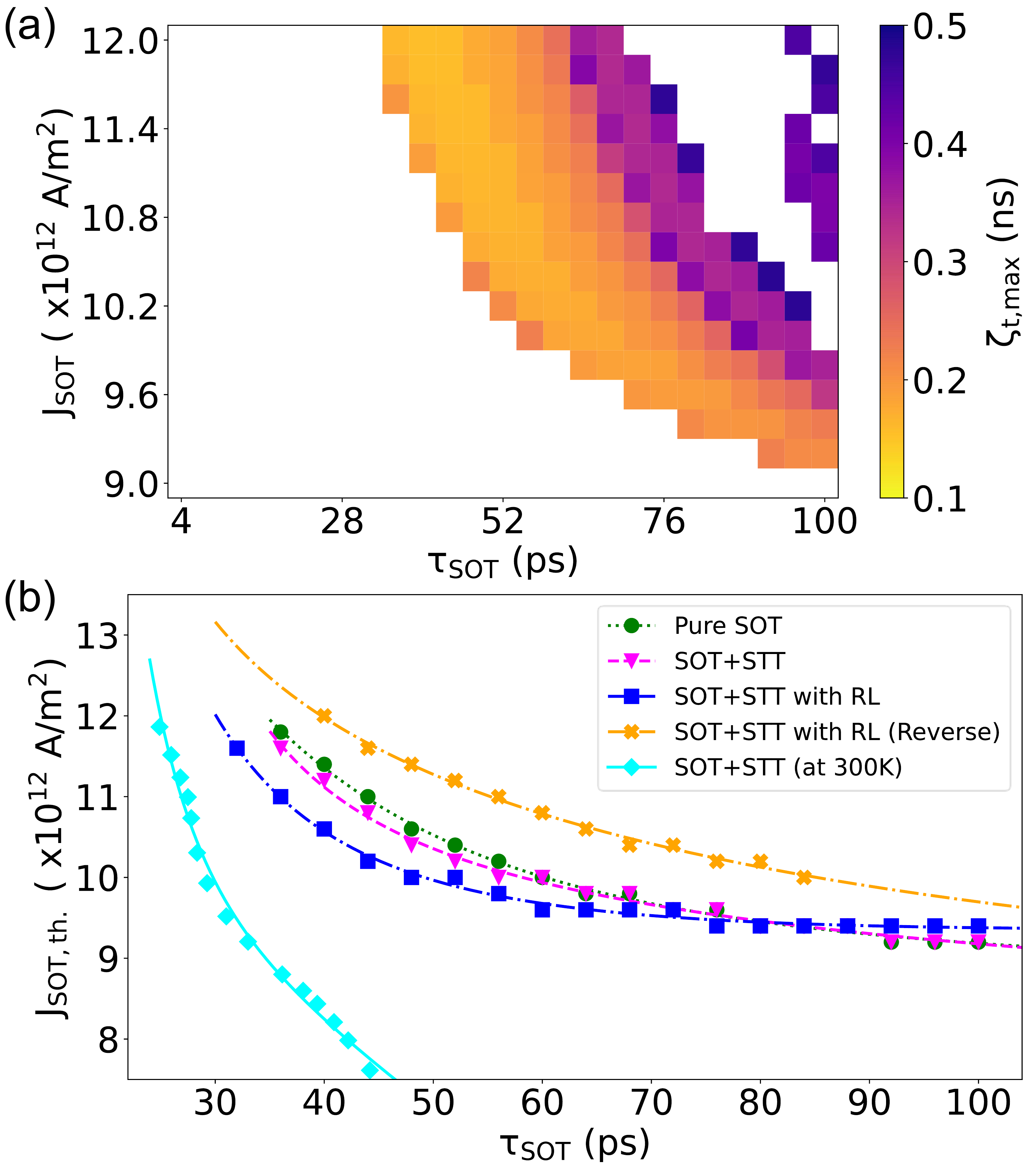}
    \caption{$\mathrm{J_{SOT}}-\mathrm{\tau_{SOT}}$ switching phasemap showing the region for $\rm AP1\rightarrow AP2$ transition for fixed $\mathrm{J_{STT}=4\times 10^{11}~A/m^2}$ and $\mathrm{\tau_{STT}=0.1~ns}$. Color encodes the maximum switching time ($\mathrm{\zeta_{t,max}}$). (b) The lower $\mathrm{J_{SOT}}-\mathrm{\tau_{SOT}}$ threshold for pure SOT and SOT+STT driven switching alongside the case for SOT+STT switching with the inclusion of stray field from reference layer (RL) derived from their individual $\mathrm{J_{SOT}}-\mathrm{\tau_{SOT}}$ switching phasemaps. The dotted curves show the polynomial fits between $\mathrm{J_{SOT}}$ and $\mathrm{1/\tau_{SOT}}$ for each case.}
    \label{fig:sotstt_map}
    \end{minipage}
\end{figure}
To optimize our protocol for energy-efficient switching, we focus on the region where switching is observed for the low values of $\mathrm{J_{SOT}}$ and $\mathrm{\tau_{SOT}}$ (dashed rectangle region in Figs.~\ref{fig:puresotpulse}(a) and (b)). For each combination of $\mathrm{J_{SOT}}-\mathrm{\tau_{SOT}}$ in this region, a fixed STT pulse of amplitude $\mathrm{J_{STT}=4\times 10^{11}\,A/m^2}$ and pulsewidth, $\mathrm{\tau_{STT}=0.1\,ns}$, is applied on FL-2 immediately after SOT ($\rm W_\tau = 0$).
The switching phasemap for this region is shown in Fig.~\ref{fig:sotstt_map}(a). The color represents the maximum switching time, $\zeta_{t,max}$, defined earlier. Since the polarity of STT current is chosen to favor $\rm AP1\rightarrow AP2$ transition, the map shows a wider region of switching compared to the pure SOT-driven switching phasemap shown in Fig.~\ref{fig:puresotpulse}. However, no significant shift in the lower threshold of the $\mathrm{J_{SOT}}-\mathrm{\tau_{SOT}}$ phasemap is observed, as this threshold represents the energy required to bring the magnetization from OOP to IP state and this energy is still provided purely by the SOT pulse. The STT pulse only helps in biasing the direction of magnetization switching. The lower $\mathrm{J_{SOT}}-\mathrm{\tau_{SOT}}$ thresholds for both pure SOT and SOT+STT cases are shown in Fig.~\ref{fig:sotstt_map}(b). Dashed curves represent the corresponding fits obtained by using a polynomial model for $\mathrm{J_{SOT}}$ as a function of $1/\mathrm{\tau_{SOT}}$. The extracted empirical relations can be further used in large-scale electrical simulators. Alongside the pure SOT and SOT+STT cases, we also show the case where the stray field of the reference layer (RL) is taken into account. With the inclusion of the RL stray field, the curve shifts favourably to the downward-left side, reducing the minimum $\mathrm{J_{SOT}}-\mathrm{\tau_{SOT}}$ needed for switching. However, it may be noted that the inclusion of the stray field introduces an asymmetry in switching thresholds between $\rm AP1\rightarrow AP2$ and $\rm AP2\rightarrow AP1$. Furthermore, the thermal statistics are obtained for the SOT phasemap (shown in Fig.~\ref{fig:sotstt_map}(a)) at 300K, and the subsequent lower switching threshold is shown in Fig.~\ref{fig:sotstt_map}(b). In the presence of thermal fluctuations, there is a significant decrease in thresholds. Detailed discussion on the inclusion of RL stray field and thermal fluctuations is given in Appendices~\ref{sec:RLeff} and ~\ref{sec:SOTmap300K} respectively.

\subsubsection{\label{sec:STTmaps}Switching Phasemaps for STT parameters}
\begin{figure*}[bthp] 
    \centering
    \includegraphics[width=\textwidth]{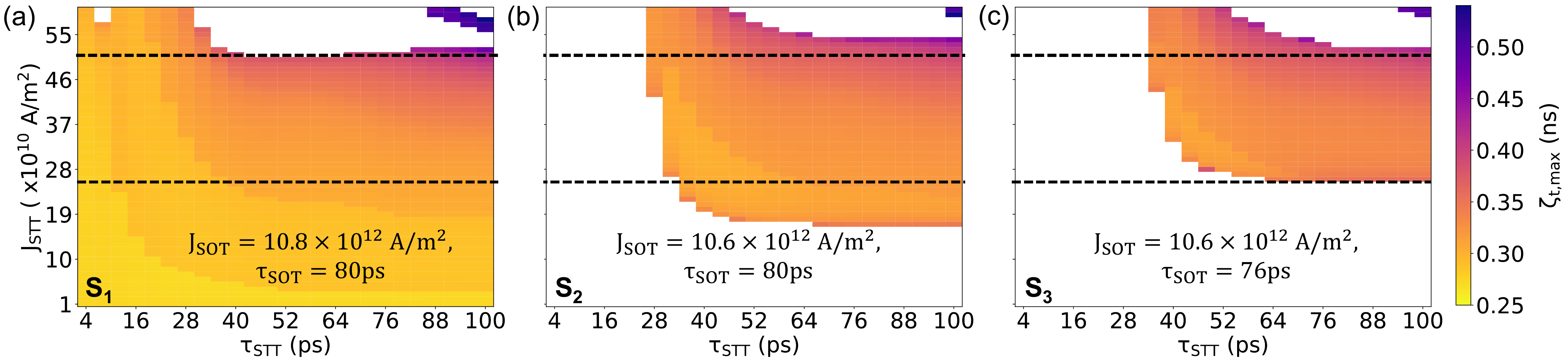} 
    \caption{Maximum switching time, $\mathrm{\zeta_{t,max}}$ phasemaps for (a) $\mathrm{J_{SOT}=10.8\times 10^{12}\,A/m^2}$ and $\mathrm{\tau_{SOT}=80ps}$, (b) $\mathrm{J_{SOT}=10.6\times 10^{12}\,A/m^2}$ and $\mathrm{\tau_{SOT}=80ps}$, (c) $\mathrm{J_{SOT}=10.6\times 10^{12}\,A/m^2}$ and $\mathrm{\tau_{SOT}=76ps}$ values.}
    \label{fig:sotstt_map_vary_STT}
\end{figure*}

In this section, we sweep the STT parameters with fixed values of SOT parameters to optimize the switching process by analyzing the $\mathrm{J_{STT}}-\mathrm{\tau_{STT}}$ phasemaps obtained for three different combinations of $\mathrm{J_{SOT}}-\mathrm{\tau_{SOT}}$ values. These values are chosen from the lowermost switching region in Fig.~\ref{fig:puresotpulse} and are marked with points S1, S2, and S3. Investigation at these points, which lie in close proximity on the pure SOT phasemap, can help in understanding the effect of small variations in the applied SOT pulse. The point S1 ($\mathrm{J_{SOT}=10.8\times 10^{12}\,A/m^2}$, $\mathrm{\tau_{SOT}=80ps}$) lies in the switching region, while points S2 ($\mathrm{J_{SOT}=10.6\times 10^{12}\,A/m^2}$, $\mathrm{\tau_{SOT}=80ps}$) and S3 ($\mathrm{J_{SOT}=10.6\times 10^{12}\,A/m^2}$, $\mathrm{\tau_{SOT}=76ps}$) lie in the region where the switching was not obtained using pure SOT. 

Figures~\ref{fig:sotstt_map_vary_STT}(a-c) show the $\mathrm{J_{STT}}-\mathrm{\tau_{STT}}$ phasemaps obtained for SOT parameters at points S1, S2 and S3. For point S1 [Figures~\ref{fig:sotstt_map_vary_STT}(a)], where switching was obtained even for the pure SOT pulse, no lower threshold for the STT pulse is seen. This is expected as the additional torque provided by STT pulse is not required if the switching can be obtained by the SOT pulse alone. However, an upper threshold for $\mathrm{J_{STT}}$, beyond which $\rm AP1\rightarrow AP2$ switching does not take place, is observed. At high $\mathrm{J_{STT}}$, the additional torque provided by the STT to FL-2 layer rotates the FL-2 magnetization significantly faster, due to which the FL-1 magnetization, which is following the FL-2 magnetization via IEC, cannot respond quickly. Due to this fast rotation of FL-2, the device undergoes $\rm AP1\rightarrow P2$ transition, which is not desirable. For points S2 and S3 [Figures~\ref{fig:sotstt_map_vary_STT}(b-c)], where the switching does not take place with SOT alone, the inclusion of STT helps with the switching process once the $\mathrm{J_{STT}}-\mathrm{\tau_{STT}}$ value crosses a threshold curve. Comparing the phasemaps for points S2 and S3, this lower threshold curve depends strongly on the chosen SOT parameters. Similar to the case of S1, an upper threshold exists for S2 and S3 as well. It may be noted that in all three phasemaps, there is an overlapping region of optimum $\mathrm{J_{STT}}$ range for which the switching happens in all three cases (dashed black lines in Fig.~\ref{fig:sotstt_map_vary_STT}(a-c)). For practical operation of the device, choosing the STT parameters within this range ensures that deterministic switching will be obtained even with small variations in the SOT pulse parameters.

\section*{\label{sec:conclusion}Conclusion}
In conclusion, we have investigated the sub-ns magnetization reversal dynamics of a Synthetic Antiferromagnet (SAF)-based p-MTJ driven by sequential sub-ns Spin-Orbit Torque (SOT) and Spin-Transfer Torque (STT) current pulses. It is shown that the SOT pulse can significantly reduce the transition time, thereby lowering the STT current threshold. 
A sequential SOT-STT-based current-pulse protocol has been designed to avoid overlap between the two current pulses while maintaining the initial and final anti-parallel states. The protocol also eliminates the need to reverse the SOT pulse polarity when transitioning from the AP-2 state back to the AP-1 state, thereby reducing the complexity of the electronic circuitry. An optimization strategy has been discussed to obtain switching for a large parameter range of SOT and STT current pulses. These results provide further insight into tuning various parameters to achieve sub-ns magnetization reversal for large variations in SOT and STT parameters, thereby enhancing the prospects for ultrafast, high-density, and robust MRAM technologies.

\section*{Data Availability}
The data supporting this study are available upon reasonable request to the authors.

\begin{acknowledgments}
{We acknowledge National Supercomputing Mission (NSM) for providing computing resources of ‘PARAM Ananta’ at IIT Gandhinagar, which is implemented by C-DAC and supported by the Ministry of Electronics and Information Technology (MeitY) and Department of Science and Technology (DST), Government of India. This work was also partially supported by the Anusandhan National Research Foundation (ANRF), India, under the project grant ANRF/IRG/2024/000636/PS. S.D. acknowledges support from the Ministry of Education (MoE), India.}
\end{acknowledgments}

\appendix
\section{\label{sec:RLeff}Switching Time Phasemaps with Reference Layer}
 
\begin{figure}[H]
    \begin{minipage}{1.0\columnwidth}
        \centering
       \includegraphics[width=\columnwidth]{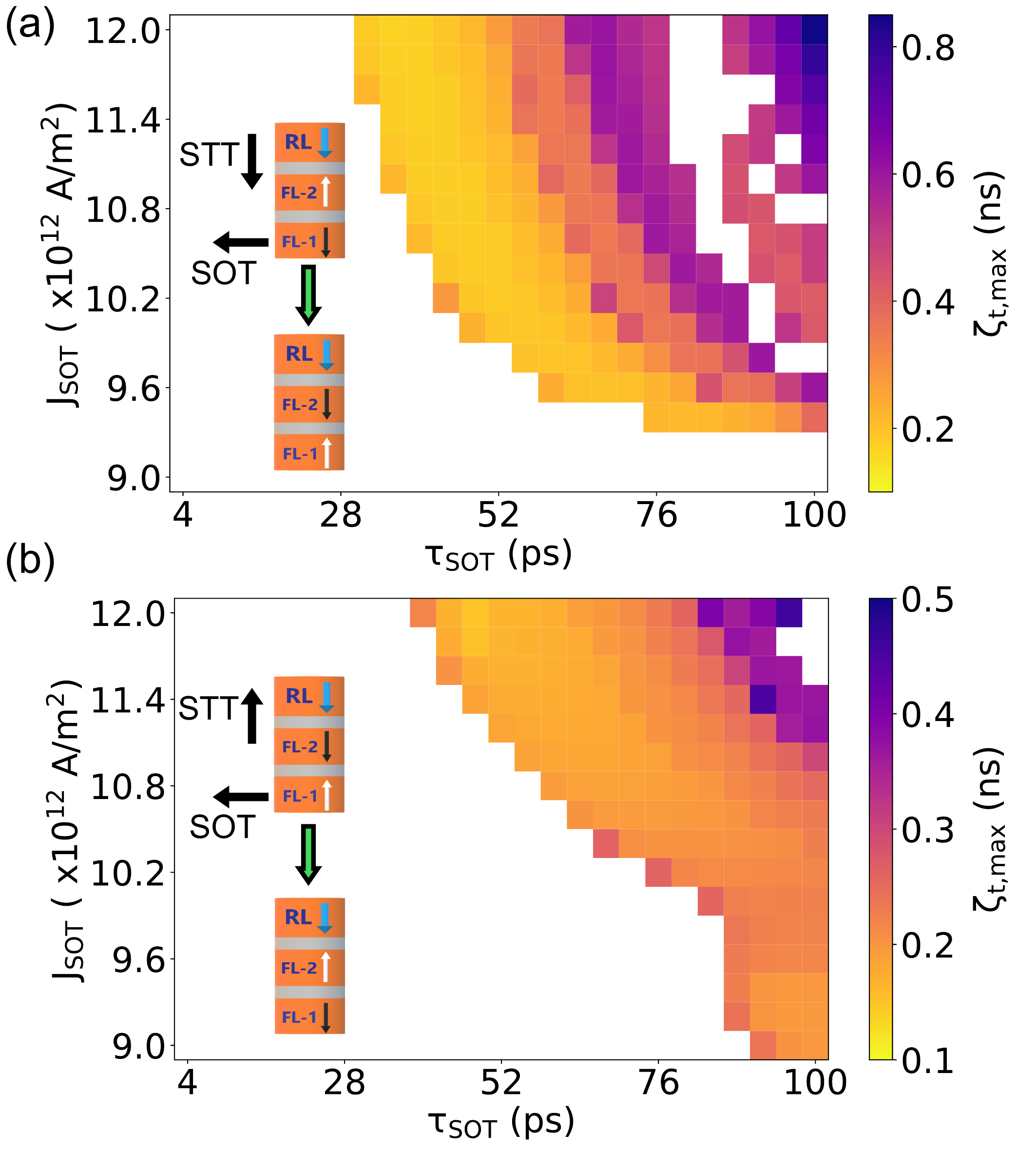}
        \caption{$\mathrm{\zeta_{t,max}}$ phasemaps for various values of SOT current amplitudes ($\mathrm{J_{SOT}}$) and pulsewidths ($\mathrm{\tau_{SOT}}$) with fixed $\mathrm{J_{STT}=4\times 10^{11}\,A/m^2, \tau_{STT}=0.1\,ns}$; (a) Switching from AP1 state to AP-2 state, and (b) Reverting back from AP-2 state to AP-1 state. (Both insets show the magnetization configurations of the initial and final states)}
        \label{fig:8withRL}
    \end{minipage}
\end{figure}

The introduction of the reference layer (RL) breaks the energy degeneracy between AP1 and AP2 states, resulting in a lower energy for the antiparallel configuration in which FL-2 is aligned parallel to RL compared to the case where FL-2 is antiparallel to RL. For the same range of $\mathrm{J_{SOT}}$, $\mathrm{\tau_{SOT}}$ values and fixed $\mathrm{J_{STT}=4\times 10^{11}\,A/m^2, \tau_{STT}=0.1\,ns}$ values as in Fig.~\ref{fig:sotstt_map}, the SOT parameter phasemap is reproduced with the addition of RL in Fig.~\ref{fig:8withRL}(a). In general, a lower threshold bound is found for $\rm AP1\rightarrow AP2$ transition compared to Fig.~\ref{fig:sotstt_map}, where the effect of RL was not included. Figure ~\ref{fig:8withRL}(b) shows a similar phasemap for the same SOT and STT parameters but for $\rm AP2\rightarrow AP1$ transition using reversed polarity of STT current. In general, a change in polarization efficiency is expected when the direction of current density is reversed in a physical MTJ stack, but it is kept the same here to isolate the effect of RL's magnetization. Comparing the lower $\mathrm{J_{SOT}}-\mathrm{\tau_{SOT}}$ threshold, switching is found to be easier for the $\rm AP1\rightarrow AP2$ compared to $\rm AP2\rightarrow AP1$ transition. This is a direct implication of the lowering of energy of the AP2 state (FL-2 parallel to RL) compared to the AP1 state (FL-2 antiparallel to RL) in the presence of RL, which makes the transition from the AP1 state to the AP2 state more favourable.

When $\mathrm{J_{SOT}=12\times 10^{12}\,A/m^2}$ is held constant and $\mathrm{\tau_{SOT}}$ is varied, as well as when $\mathrm{\tau_{SOT}=100\,ps}$ is fixed and $\mathrm{J_{SOT}}$ is varied, the widths of the switching regions are found to be greater in the former scenario and reduced in the latter (Fig.~\ref{fig:8withRL}(a)), in comparison to those reported in Fig.~\ref{fig:sotstt_map}. RL has polarization along $-\hat{z}$, so the stray fields are also along $-\hat{z}$ for both FLs. As FL-1 is trying to align in-plane, it is countered by the stray fields of RL. Therefore, even when applying continuous SOT, FL-1 has a slightly negative $\mathrm{m_z}$ component, and through IEC, FL-2 has a slightly positive $\mathrm{m_z}$ component. Additionally, the damped precessions begin earlier for the case with RL, and the $\mathrm{m_z}$ of FL-2 damps more rapidly. Therefore, the $\mathrm{m_z}$ precesses closer to the IP region for a longer time. Hence, the range of switching for pulse width with fixed amplitude is larger with RL. For low SOT amplitudes and a fixed pulse width of 0.1 ns, increasing the amplitude decreases the number of precessions. Although with RL, the $\mathrm{m_z}$ of FL-2 undergoes more precessions (increase in frequency of oscillations seen while applying a field orthogonal to the current polarization\cite{nanooscill}). The combined effect further reduces the number of precessions; consequently, the switching amplitude range for a fixed pulse width is broader when using RL.

\section{\label{sec:SOTmap300K}Temperature Statistics for SOT parameters}

\begin{figure}[H]
    \begin{minipage}{1.0\columnwidth}
        \centering
       \includegraphics[width=\columnwidth]{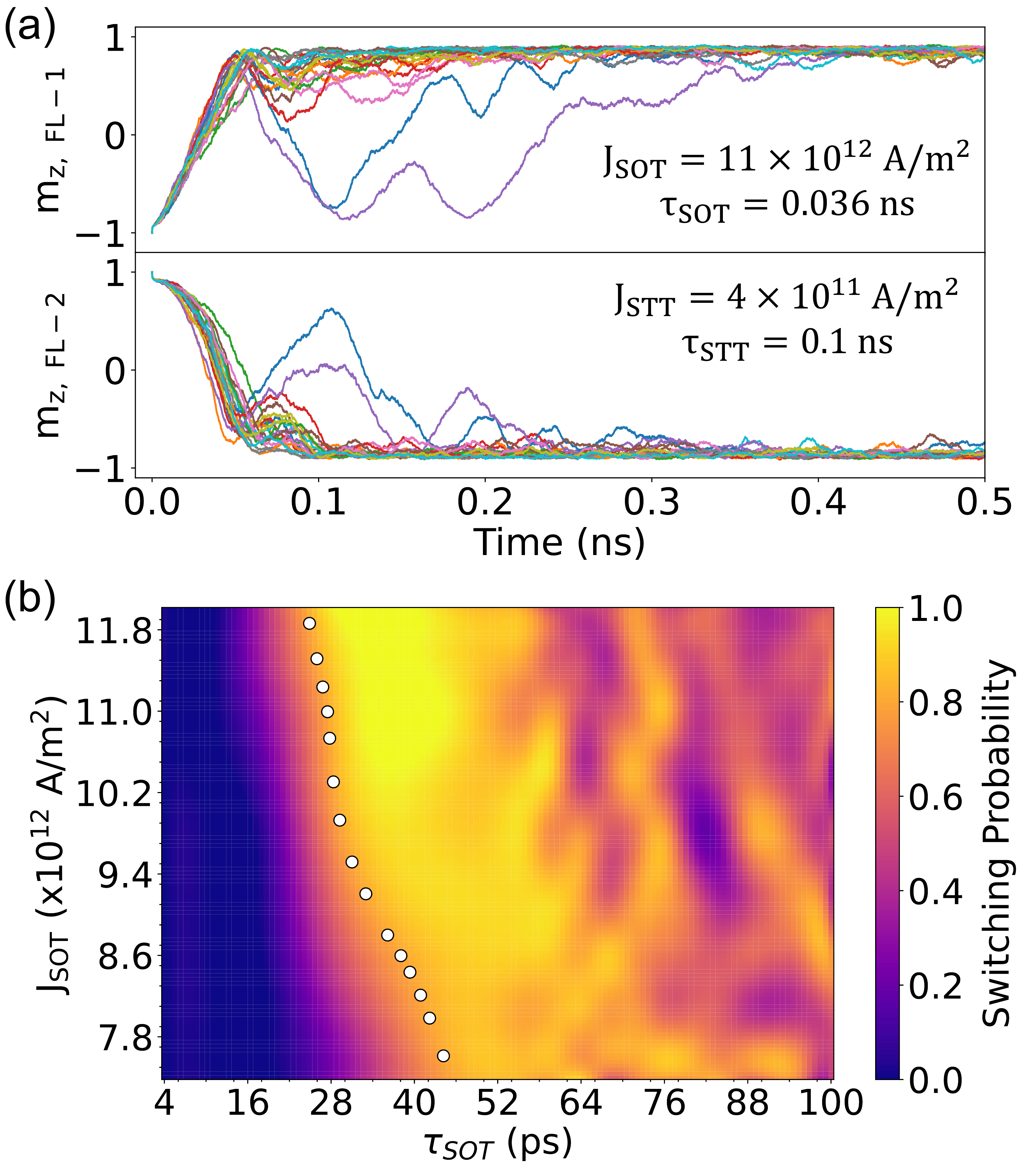}
        \caption{(a) $\mathrm{m_z}$ dynamics of FL-1 and FL-2 at 300K for $\mathrm{J_{SOT}=11\times 10^{12}\,A/m^2, \tau_{SOT}=0.036\,ns}$ and $\mathrm{J_{STT}=4\times 10^{11}\,A/m^2, \tau_{STT}=0.1\,ns}$ values; (b) Switching probability phasemap for various values of SOT current amplitudes ($\mathrm{J_{SOT}}$) and pulsewidths ($\mathrm{\tau_{SOT}}$) with fixed $\mathrm{J_{STT}=4\times 10^{11}\,A/m^2, \tau_{STT}=0.1\,ns}$. The white dots indicate a switching probability of 0.8.}
        \label{fig:9Thermal}
    \end{minipage}
\end{figure}

Fig.~\ref{fig:9Thermal}(a) shows the $\mathrm{m_z}$ switching dynamics at different thermal seeds (T=300K) for $\mathrm{J_{SOT}=11\times 10^{12}\,A/m^2, ~\tau_{SOT}=0.036\,ns}$ and, $\mathrm{J_{STT}=4\times 10^{11}\,A/m^2, ~\tau_{STT}=0.1\,ns}$ values. Fig.~\ref{fig:9Thermal}(b) shows the full phasemap for switching probability calculated for different $\mathrm{J_{STT}-\tau_{STT}}$ values. As discussed previously, there is a significant decrease in the SOT pulsewidth threshold as current density increases.

\section{\label{sec:analytSTTcont}Derivation of Threshold Current Density for Continuous STT-Driven Switching}
Switching with only continuous STT current in FL-2 is discussed in Sec.~\ref{sec:pureSTTall}. In this section we have analytically derived the threshold current density required for switching using the stereographic projection method\cite{SGPKomineas}. First, we can introduce two complex variables from the six magnetization components as follows.
\begin{equation}
    \Omega = \frac{(m_{x,2}+\iota m_{y,2})}{(1+m_{z,2})}
\end{equation}
\begin{equation}
    \psi = \frac{(m_{x,1}+\iota m_{y,1})}{(1-m_{z,1})}
\end{equation}

Applying stereographic projection formalism in equations \ref{eq:LLG1} and \ref{eq:LLG2}, we obtain the following coupled equations.

\begin{align}
    &(\iota + \alpha)\frac{d\Omega}{dt}=-\frac{\gamma (\Omega \bar{\Omega}+1)^2}{2} \frac{\partial E_{total}}{\partial \bar{\Omega}}  \nonumber \\
&\quad - \frac{\iota \gamma \beta_{ST}}{2} [\sigma_x(\Omega^2 -1)-\iota \sigma_y(\Omega^2 +1)+2\sigma_z \Omega]
    \label{eq:sterST1}
\end{align}
\begin{align}
    &(\iota - \alpha)\frac{d\psi}{dt}=\frac{\gamma (\psi \bar{\psi}+1)^2}{2} \frac{\partial E_{total}}{\partial \bar{\psi}}  \nonumber \\
&\quad - \frac{\iota \gamma \beta_{ST}}{2} [\sigma_x(\psi^2 -1)-\iota \sigma_y(\psi^2 +1) -2\sigma_z \psi]
    \label{eq:sterST2}
\end{align}

where, 
\begin{align}
E_{total}=2|J_{ex}|\hat{m}_1\cdot \hat{m}_2 ~-\frac{\kappa}{2}[m_{1,z}^{2}+m_{2,z}^{2}]
\label{eq:etotster}
\end{align}
\begin{align*}
    E_{total}&=2|J_{ex}|\hat{m}_1\cdot \hat{m}_2 ~-\frac{\kappa}{2}[m_{1,z}^{2}+m_{2,z}^{2}] \\
    \beta_{STT}&=\frac{J_z \hbar P_{STT}}{2M_S e d} \\
    \quad \kappa&=\frac{2K_{eff}}{M_{S}}=\frac{2}{M_{S}}(K_{u_1}-\frac{1}{2} \mu_0 M_{S}^{2})
\end{align*}

$E_{total}$ represents the total energy of the coupled system. The first term represents inter-layer exchange energy term. The second term is a combination of uniaxial anisotropy and demagnetization energy terms. After applying STT of -z polarization in FL-2, and simplifying Eq.~\ref{eq:sterST1},~\ref{eq:sterST2} we obtain the following coupled equations.
\begin{align}
    (\iota + \alpha)\frac{d\Omega}{dt}&=\frac{2 \gamma |J_{ex}|}{(\psi \bar{\psi}+1)}(\Omega^2 \bar{\psi} - \psi - \Omega + \Omega \psi \bar{\psi})  \nonumber \\
&\quad  + \kappa \gamma\frac{\Omega (\Omega \bar{\Omega}-1)}{(\Omega \bar{\Omega}+1)}+\iota \gamma \beta_{STT} \Omega
    \label{eq:sterSTT1}
\end{align}
\begin{align}
    (\iota - \alpha)\frac{d\psi}{dt}&=-\frac{2 \gamma |J_{ex}|}{(\Omega \bar{\Omega}+1)}(\psi^2 \bar{\Omega} - \psi - \Omega + \psi \Omega \bar{\Omega})  \nonumber \\
&\quad  - \kappa \gamma\frac{\psi (\psi \bar{\psi}-1)}{(\psi \bar{\psi}+1)}
    \label{eq:sterSTT2}
\end{align}

We have observed in our simulations that only $\mathrm{\hat{m}_2}$ will switch. So, we adopt the following ansatz.
\begin{equation*}
    \Omega (t)=Ae^{(B+\iota \omega)t}~;~~\psi(t)=\frac{1}{A} e^{(B+\iota \omega)t} e^{\iota \phi}
\end{equation*}
Substituting $\mathrm{\Omega}$ and $\mathrm{\psi}$ in Eq.~\ref{eq:sterSTT1}, ~\ref{eq:sterSTT2} and simplifying we obtain the following equations.
\begin{widetext}
\begin{align}
    (\iota+\alpha)(B+\iota \omega)=\gamma \Bigg[ -\frac{2|J_{ex}| e^{-\iota \phi}(A^2+e^{\iota \phi})(-e^{2Bt}+e^{\iota \phi})}{(A^2+e^{2Bt})} + \frac{\kappa (-1+A^2 e^{2Bt})}{(1+A^2 e^{2Bt})} +\iota \beta_{STT} \Bigg]
    \label{eq:eq1sim}
\end{align}
\begin{align}
    (-\iota+\alpha)(B+\iota \omega)=\gamma \Bigg[ \frac{2|J_{ex}|e^{-\iota \phi}(-1+e^{2Bt+\iota \phi})(A^2+e^{\iota \phi})}{(1+A^2 e^{2Bt})} + \frac{\kappa (-A^2 + e^{2Bt})}{(A^2 + e^{2Bt})} \Bigg]
    \label{eq:eq2sim}
\end{align}
\end{widetext}
Adding and subtracting Eq.~\ref{eq:eq1sim} and ~\ref{eq:eq2sim}, we obtain two new equations. After expanding the complex terms in those equations and extracting the real and imaginary coefficients, we obtain the four coupled equations shown below.
\begin{widetext}
\begin{align}
    B \alpha ( 1+A^4) e^{2Bt} + A^2 [B \alpha (1+e^{4Bt}) - \gamma (2|J_{ex}|+\kappa) (-1+e^{4Bt})]=0
    \label{eq:simp1}
\end{align}
\begin{align}
    (\gamma \beta_{STT}  - 2\alpha \omega)[e^{2Bt} + A^4 e^{2Bt}+ A^2(1+e^{4Bt})] =2 \gamma |J_{ex}|(-1+A^4)(-1+e^{4Bt}) sin(\phi)
    \label{eq:simp2}
\end{align}
\begin{align}
A^2 \omega (1+e^{4Bt}) + e^{2Bt} [-2\gamma |J_{ex}|(-1+A^4)+\gamma \kappa(-1+A^4) +\omega(1+A^4)] +\gamma |J_{ex}|(-1+A^4)(1+e^{4Bt})cos(\phi) = 0
\label{eq:simp3}
\end{align}
\begin{align}
(2B-\gamma \beta_{STT} )[e^{2Bt}+A^4 e^{2Bt}+A^2(1+e^{4Bt})] +2 \gamma |J_{ex}|[1+4A^2 e^{2Bt}+e^{4Bt}+A^4 (1+e^{4Bt})]sin(\phi)=0
\label{eq:simp4}
\end{align}
\end{widetext}
As STT is being applied continuously, the above four equations should hold every time. So, for Eq.~\ref{eq:simp1}, the coefficients of the exponential terms should be individually zero and from the coefficients of $e^{2Bt}$ we get,
\begin{equation*}
    \alpha B (1+A^4)=0
\end{equation*}
We know that, $\alpha \ne 0;~A\ne0;~(1+A^4)\ge1>0$, which yield $B=0$. This means our solution is oscillatory without any damping. Substituting the value of B in Eq.~\ref{eq:simp2}, yields;
\begin{align*}&
(\gamma \beta_{STT}  - 2\alpha \omega)(1+A^2)^2=0;~ \nonumber \\
&\quad \implies (\gamma \beta_{STT}  - 2\alpha \omega)=0\implies \omega=\frac{\gamma \beta_{STT}}{2 \alpha}
\end{align*}
Again substituting the value of B in Eq.~\ref{eq:simp4}, yields;
\begin{align*}&
    \gamma (1+A^2)^2[-\beta_{STT} + 4|J_{ex}|sin(\phi)] =0 \nonumber \\
&\quad \implies \beta_{STT}=4 |J_{ex}|sin(\phi)  \nonumber \\
&\quad \implies \phi=sin^{-1} \Big(\frac{\beta_{STT}}{4 |J_{ex}|} \Big)~;  \nonumber \\
&\quad \mathrm{and,}~cos(\phi)=\pm \sqrt{1-\Big(\frac{\beta_{STT}}{4 |J_{ex}|}\Big)^2}
\end{align*}
Similarly from Eq.~\ref{eq:simp3}, we can extract the relation for A, which provides;
\begin{equation*}
    A=\pm \sqrt{\frac{2\alpha \kappa_r + \beta_{STT}}{2\alpha \kappa_r - \beta_{STT}}};~ \mathrm{where,}~ \kappa_r=2 |J_{ex}|[1-cos(\phi)]-\kappa
\end{equation*}
We know that $\Omega$ diverges if $\mathrm{\hat{m}_2}$ switches from +z to -z direction. If we observe the analytical solution, $\Omega$ diverges when;
\begin{equation*}
    2\alpha \kappa_r = \beta_{STT}
\end{equation*}
Finally we obtain the relation for the critical current density,
\begin{widetext}
\begin{align}&
J_{STT,0}^{analytical}=\frac{2 e \alpha d}{P_{STT} \hbar (1+\alpha^2)} \Bigg[-4 K_{u_{1}}+ 2 M_S(2|J_{ex}|+\mu_0 M_S) \nonumber \\
&\quad +\sqrt{-4\alpha^2 K_{u_{1}}^2+4J_{ex}^2 M_S^2 +4\alpha^2K_{u_{1}}M_S(2|J_{ex}|+\mu_0 M_S)- M_S^3 \mu_0(4|J_{ex}|+\mu_0 M_S)} \Bigg]
    \label{eq:critcd}
\end{align}    
\end{widetext}
After substituting the values of the material parameters in Eq.~\ref{eq:critcd}, we obtain the threshold STT current density required for switching to be $\sim4.18\times10^{11}~A/m^2$. The obtained value have the same order as the value obtained from extrapolating the linear fit in Fig.~\ref{fig:2abc}(b), indicating the model provides a reliable prediction of the reversal dynamics.

\bibliography{references}

\end{document}